# Analysis of the Gas states at Liquid/Solid Interface based on interactions at the microscopic level

*Zhaoxia Li,[†,‡] Xuehua Zhang,[§] Lijuan Zhang,[†,‡] Xiaocheng Zeng,[∥] Jun Hu,[*,†,§] and Haiping Fang[*,†]*

Shanghai Institute of Applied Physics, Chinese Academy of Sciences, P.O. Box 800-204, Shanghai 201800, China

Graduate School of the Chinese Academy of Sciences, Beijing 100080, China

Bio-X Life Sciences Research Center, College of Life Science and Biotechnology, Shanghai JiaoTong University, Shanghai 200030, China

Department of Chemistry and Center for Materials Research & Analysis, University of Nebraska-Lincoln, Lincoln, NE 68588, USA

* Corresponding authors. Email: fanghaiping@sinap.ac.cn, jhu@sjtu.edu.cn.

**RECEIVED DATE (to be automatically inserted after your manuscript is accepted if required according to the journal that you are submitting your paper to)**

The states of gas accumulated at the liquid/solid interface are analyzed based on the continuum theory where the Hamaker constant is used to describe the long-range interaction at the microscopic scale. The Hamaker constant is always negative, whereas the "gas" spreading coefficient can be either sign. Despite the complexity of gas, including that the density profile may not be uniform due to absorption on both solid and liquid surfaces, we predict three possible gas states at the liquid/solid interface, i.e. complete "wetting", partial "wetting" and pseudopartial "wetting". These possible gas states correspond

---

[†] Shanghai Institute of Applied Physics, Chinese Academy of Sciences

[‡] Graduate School of the Chinese Academy of Sciences

[§] Shanghai JiaoTong University

[∥] University of Nebraska-Lincoln

respectively to a gas pancake (or film) surrounded by a wet solid, a gas bubble with a finite contact angle, and a gas bubble(s) coexisting with a gas pancake. Typical thickness of the gas pancakes is at the nanoscale within the force range of the long-range interaction, whereas the radius of the gas bubbles can be large. The state of gas bubble(s) coexisting with a gas film is predicted theoretically for the first time. Our theoretical results can contribute to the development of a unified picture of gas nucleation at the liquid/solid interface.

**Introduction**

When gas on the nanoscale nucleates at the liquid/solid interface, the interfacial properties and dynamics of the system can be significantly changed, for example, by the occurrence of a long-range interaction between hydrophobic interfaces immersed in water,[1,2] a long slip length in hydrodynamics[3-5] and the reduction of friction in microfluidic transportation.[3,6,7] The existence of sub-microscopic gas bubbles on solid surfaces immersed in water was first proposed to account for the unexpected long-ranged attractive force between hydrophobic surfaces.[1,2] Since then, a range of experimental evidence has been presented in support of the existence of nanoscale gas nucleated at the water/hydrophobe interface,[8-19] including direct imaging of nanobubbles by tapping mode atomic force microscopy (AFM),[10-16] and the density deficit of a thin water layer at water/hydrophobe interfaces measured by neutron and X-ray reflectivity.[17-19]

New theories have also been developed on this aspect. Lum, Chandler and Weeks have developed a theory of hydrophobicity and predicted the existence of a gas-like layer between two extended hydrophobic planar surfaces in water.[20] De Gennes has shown that a very large slip length can result from a thin gas layer.[3] Molecular dynamics simulations have also been used to study the gas and dewetting behavior between two hydrophobic planar surfaces and even in a two-domain protein immersed in water.[21-23]

However, whether gas that accumulates on the nanoscale can be stable at the liquid/solid interface is still a debatable question.[24-35] The pressure inside a gas bubble with a radius at nanometer scale is extremely high based on the traditional Laplace-Young equation.[24] The lifetime of those nanobubbles predicted by conventional thermodynamics only ranges from several picoseconds to hundreds of microseconds,[25] which is too short to be experimentally reproducible (depending on the technique). On the other hand, many experimental results on gas accumulation have been in conflict, which likely arises from the influence of many factors such as complicated behavior of hydrophobic self-assembled monolayer surfaces,[26-28] possible experimental artifacts,[29,30] or dissolved gases.[31-35] Recently, the debate has been focused on whether the state of gas at the liquid/solid interface is a gas bubble or a gas layer. Steitz et al. suggested that a precursor gas layer contributed to the depletion layer of D2O, and nanobubbles were induced by the tip of AFM.[19] In 2005, Doshi et al. pointed out that the density reduction of water at the hydrophobic silane-water interface depended on the dissolved gases in water, whereas preexisting nanobubbles were excluded.[31] However, lately, Attard et al. insisted that nanobubbles were responsible for the long-range attractive force measured by AFM between hydrophobic surfaces immersed in water.[35] Very recently, by x-ray reflectivity measurement and using of degassed water, it was shown that the depletion layer between the water and a hydrophobic surface cannot be explained by nanobubbles.[36] On the other hand, Zhang et al. have reported that the gas nanobubbles[11, 14, 37] and gas pancakes[16] can be formed with high reproducibility by the solvent exchange method.

We hold the view that the debates or the conflict between theory and experiments may be due to lack of knowledge of nanoscale gas phases at the water/solid interface. A similar situation occurred regarding liquid nucleation more than 20 years ago. Through much careful effort by many researchers, the physicochemical parameters controlling the conventional thermodynamic wettability of solid surfaces were clarified in 1970s.[38] In 1980s, the deviations from thermodynamic equilibrium were about to be understood and it was realized that long-range forces (van der Waals, electrostatic, etc.) at the microscopic level are essential.[38, 39] De Gennes and coworkers were the first to put forward a unified picture of liquid nucleation with a continuum theory incorporating interaction potentials at the microscopic level.[38, 40] That theory predicts three states of liquid on the solid surface: complete wetting, partial wetting and pseudopartial wetting, corresponding to liquid pancake (film), liquid droplet surrounded by a dry solid, and liquid droplet coexisting with a liquid film. The conditions for the existence of these three states and the thickness of pancakes depend on the interaction potentials and the volume of the liquid accumulated.[38, 40] Remarkably, these theoretically predicted states have since been observed in experiments,[41-44] showing the robustness of the theory.

Obviously, the interactions at the microscopic scale should also be an indispensable ingredient in the theory of gas nucleation. We expect that the seminal theory of liquid wetting developed by de Gennes and coworkers can be extended to the study of gas nucleation at the liquid/solid interface. It should be noted that the analysis is even more complex due to the particularity of gas. For example, the density of gas accumulated at the liquid/solid interface should vary with the long- and short-range interactions between gas, liquid and solid molecules at the microscopic level, and is considerably smaller than those of ambient liquid and solid. By carefully taking into account the particularity of gas, we will show in this Letter that gas layers and gas bubbles are both possible states of gas nucleated at the liquid/solid interface. Moreover, a new state, a gas bubble(s) coexisting with a gas film with a finite thickness, is predicted by the theory.

**Theoretical Analysis**

Following the pioneering idea of Brochard-Wyart et al.,[40] the thermodynamic free energy $f(e)$ (per unit area) of a uniform gas film of thickness $e$ between an ideal smooth solid surface and a liquid surface reads

$$f(e) = \gamma_{sg} + \gamma_{lg} + P(e), \quad (1)$$

where $\gamma_{sg}$ ($\gamma_{lg}$) is the solid/gas (liquid/gas) interfacial tension, and $P(e)$ is the interaction potential per unit area between the liquid and solid with gas in-between at the microscopic level.

When $e$ is much larger than typical molecular size $a_0$, $P(e)$ is controlled by the long-range retarded van der Waals forces,[40]

$$P(e) = \frac{A}{12\pi e^2}, \quad a_0 \ll e \ll l, \quad (2)$$

where $l$ is an ultraviolet cutoff length, and $A$ is the effective Hamaker constant given by $A = -K(\alpha_l - \alpha_g)(\alpha_s - \alpha_g)$.[40, 45] Here $\alpha_l$, $\alpha_g$ and $\alpha_s$ are the polarizabilities per unit volume of the liquid, gas and solid, respectively, and $K$ is a positive constant (see Appendix of Ref. [40]). The density of gas is usually much less than that of solid or liquid. Hence, $\alpha_g$ is always small compared to $\alpha_l$ and $\alpha_s$, and thus the Hamaker constant $A$ is always negative.

For small film thicknesses, $P(e)$ depends on the detailed short-range interactions but approaches

$$P(e \to 0) = S_g = \gamma_{sl} - \gamma_{sg} - \gamma_{lg}, \quad (3)$$

where $S_g$ is introduced as the "gas spreading coefficient" and $\gamma_{sl}$ is the liquid/solid interfacial tension. If only the attractive part of the van der Waals interaction is taken into account, $S_g$ has the same sign as $A$ which is negative.[38, 40] However, for a small thickness $e$, there exist certain short-range interactions such as the repulsive part of the van der Waals interaction. There are still possibilities for a positive gas spreading coefficient $S_g$.

We now present the conditions for the final equilibrium states of gas nucleated at a liquid/solid interface. The free energy of a volume of gas with a coverage area $\mathcal{A}$ and height $e$ is[40]

$$F(e) = F_0 - \mathcal{A}S_g + P(e)\mathcal{A}, \quad (4)$$

where $F_0$ is a constant. In contrast to the case of liquid nucleation, the density of the gas at the liquid/solid interface can be easily changed. For example, the density can vary with the height of gas due to the van der Waals force and other possible forces at the microscopic level. The conservation of volume in the liquid case should be replaced by the conservation of the total number of gas molecules here. This molecular number conservation is based on the observed experimental evidences. The experiments show that the morphology of nanobubbles can remains almost unchanged at least over hours.[16] For the gas pancake, about 1.5 hours after the formation of pancakes at the temperature of 25°C, the morphology of pancakes does not change for a very long time.[16] Since the gas is absorbed by both sides (solid and liquid), the density of the gas also depends on the thickness $e$ of the gas layer. The total number of gas molecules accumulated is

$$N = \mathcal{A}(e)\int_0^e \rho(x,e)dx, \qquad (5)$$

Where $\rho(x,e)$ is the gas density with a distance $x$ above the solid surface in a gas layer with a thickness $e$. Taking the minimum of the total free energy of the gas with the constraint that the total number of gas molecules stays fixed leads to

$$f(e) = \gamma_{sl} - \Pi(e)\frac{\int_0^e \rho(x,e)\cdot dx}{\rho(e) + \int_0^e \frac{\partial \rho(x,e)}{\partial e}\cdot dx}, \qquad (6)$$

where $\Pi(e) = -df(e)/de = -dP(e)/de$ is the disjoining pressure[40] and the dependence of both $\gamma_{sg}$ and $\gamma_{lg}$ on the density of gas is small because the density of gas is usually small. The small modifications of $\gamma_{sg}$ and $\gamma_{lg}$ due to the density change of gas do not change the conclusions of this paper. If there is a solution $e_s$, for this equation, a gas pancake with an equilibrium thickness $e_s$ is conceivable. In the case of a uniform density, this equation simplifies to that for the liquid case:

$$f(e) = \gamma_{sl} - \Pi(e)\cdot e \qquad (7)$$

The energy function $f(e)$ usually oscillates at atomic scales due to the complex interactions between the molecules. Following the original idea of Brochard-Wyart et al.,[40] we assume that the free energy function $f(e)$ interpolates smoothly between the long-range van der Waals tail and the wet limit ($P(0) = S_g$). The oscillation of $f(e)$ has no effect on the conclusion in this paper for $A < 0$.[40] In this gas case, $\Pi(e) > 0$ for $e < e_m$, where $e_m$ is the first position of the minimum of $f(e)$.

**Results and Discussion**

Now we discuss the possible states of gas nucleated at the liquid/solid interface from eqs 4-6 for $S_g > 0$ and $S_g < 0$, respectively.

(i) When $S_g > 0$ and $A < 0$, non-zero contact angle for a gas bubble does not exist because $(\gamma_{sl} - \gamma_{sg})/\gamma_{lg} > 1$. In this gas case, the analysis of whether there is a non-zero solution of eq 6, containing both the differential and the integral of the density profile, is usually very complex. Fortunately, we can prove that the condition for the existence of gas pancakes at the liquid/solid interface is independent of the non-uniformity of the gas density profile. Explicitly, we can prove that if eq 7 has a solution in the interval of $0 < e < e_m$, there must be a solution for the eq 6. We first define

$$G(e) = f(e) - [\gamma_{sl} - \Pi(e) \cdot e], \tag{8}$$

$$G'(e) = f(e) - [\gamma_{sl} - \Pi(e) \frac{\int_0^e \rho(x,e) \cdot dx}{\rho(e) + \int_0^e \frac{\partial \rho(x,e)}{\partial e} \cdot dx}]. \tag{9}$$

These two functions are constructed such that they vanish for values of $e$ that satisfy eq 7 and eq 6 respectively. From eq 1 and eq 3, $G(e \to 0) = G'(e \to 0) = 0$ - i.e. the two functions have the same value at $e = 0$. We also have that $G(e_m) = G'(e_m)$, since $\Pi(e) = 0$ at $e = e_m$. Furthermore, for small but non-zero values of $e$ (smaller than the decay length of the gas density profile) the two functions must have the same sign; this is because the gas density profile must get closer and closer to uniformity as $e$ approaches zero, and so $G(e)$ and $G'(e)$ must converge to the same trend in this limit. Now, if eq 7 has a solution $e_s$ in the interval $0 < e < e_m$, then $G(e)$ changes signs between $e = 0$ and $e = e_m$, and so must $G'(e)$; i.e. there is a value of $e_s$ in the interval of $0 < e < e_m$ at which $G'(e_s) = 0$. Consequently, for the potential function $f(e)$, a gas pancake with a finite thickness is always conceivable no matter how the density varies, provided there is a solution for eq 7 corresponding to a uniform gas density. Thus, testing for the existence of the possible states of gas nucleated at the liquid/solid interface can be simplified by assuming a uniform gas density and then proceeding in the same way as that proposed by Brochard-Wyart et al..[40] However, the precise thickness of a gas pancake depends on the explicit gas density profile.

The final state of the gas film depends on the number $N$ of gas molecules nucleated at the liquid/ solid interface. When there exists a solution of eq 6 and the gas nucleated has a number $N$ of molecules with $N < \mathcal{A}_s \int_0^{e_s} \rho(e, e_s) de$, where $\mathcal{A}_s$ is the total area of the perfect solid surface, there is a gas pancake with gas coverage $N / \int_0^{e_s} \rho(e, e_s) de < \mathcal{A}_s$ as shown in Figure 1(a). If the number of gas molecules nucleated falls in the interval $\mathcal{A}_s \int_0^{e_s} \rho(e, e_s) de < N < \mathcal{A}_s \int_0^{e_m} \rho(e, e_m) de$, a gas film will cover the ideal surface with a thickness $e_s \leq e_m$ since the thicker the gas film, the smaller the free energy $F(e)$. In the case that there is no solution of eq 6, the final state corresponds to a zero-thickness film, $e \to 0$, when the gas molecular number $N < \int_0^{e_m} \mathcal{A}_s \rho(e, e_m) de$. When $N > \int_0^{e_m} \mathcal{A}_s \rho(e, e_m) de$, the free energy of a gas film with height $e_m$ covering the whole solid surface is minimal for both cases since $e = e_m$ is the position of the minimum of $f(e)$. The state of a simple film with a larger thickness $e > e_m$ is unstable. Therefore, the additional gas molecules may exist in the form of gas bubble, which is located on the top of the gas film. The contact angle $\theta_m$ for the gas bubble is defined by $\cos\theta_m = 1 + P(e_m)/\gamma_{lg}$.[40] The final equilibrium state may thus be pseudopartial "wetting", that is a bubble coexisting with a gas film of thickness $e_m$, as displayed in Figure 2(b).

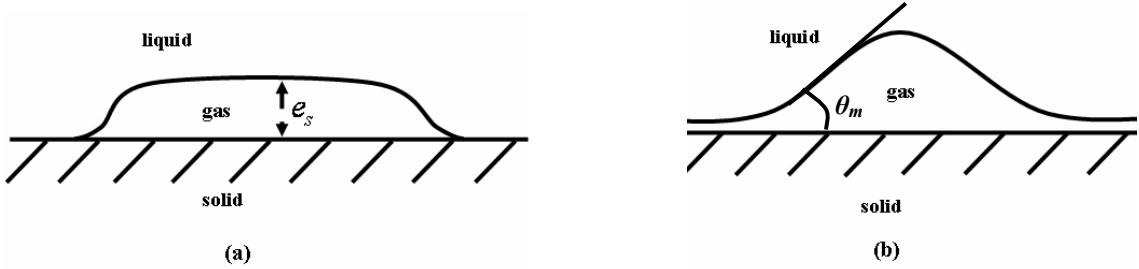

**Figure 1.** A schematic map showing the states of gas at the liquid/solid interface. (a) A gas pancake with a thickness $e_s$. (b) A gas bubble coexisting with a gas film of thickness $e_m$ at contact angle $\theta_m$.

(ii) $S_g < 0$ and $A < 0$. There are three possible cases for different kinds of function $f(e)$. The first two cases are the same as those for $S_g > 0$ and $A < 0$ except that $\gamma_{sl}$, the intercept of the curve $f(e)$ with the vertical axis, is less than $(\gamma_{sg} + \gamma_{lg})$. In the third case, $f(e)$ increases monotonically, and there exists a non-zero gas contact angle $\theta_g$ as given by the Young equation:

$$\gamma_{lg} \cos\theta_g = \gamma_{sl} - \gamma_{sg}. \tag{10}$$

The final equilibrium state may be a gas bubble.

The stable existence of nanoscale gas bubbles at the liquid/solid interface has been supported by a range of experimental evidence.[10-15] Very recently, all the three possible interfacial gaseous states were observed at the water/HOPG system under different conditions.[16] Here we show some of the primary understanding about the comparison between the theoretical analysis and the experimental results.

In our theory, the different formation conditions of three gaseous states may correspond to the different free energy of the system. As discussed above, the morphology of the gaseous states is closely related to the number of gas molecules available on the surface. If eq 6 has a solution, a small quantity of gas molecules will aggregate in the form of gas pancake. With the increase of the gas molecules, when the number of the gas exceeds $\int_0^{e_m} A_s \rho(e, e_m) de$, the state of nanobubble on gas pancake is formed, where the thickness of the gas pancake is $e_m$.

Zhang has reported experimentally that the size of micro-pancakes and nanobubble-on-pancakes can be quite sensitive to certain external factors, for example, the change of temperature, and the addition of ethanol in water.[16] The increase of the temperature can induce the augment of the coverage of the gas pancakes.[16] The theoretical analyses in our manuscript show that if the equilibrium thickness $e_s$ changes, the coverage of the pancakes will change too by assuming that change of temperature does not alter the whole gas molecular number. The size of micro-pancakes may be enlarged by the increase of temperature through diminishing the equilibrium thickness $e_s$. When the concentration of ethanol in water is greater than 5% (Vol.), the micropancake state becomes unstable and eventually vanishes.[16] All micropancakes in water have diminished after the displacement of water with 10% (Vol.) ethanol solution.[16] Theoretically, the increase of the concentration of ethanol aqueous solution may change the effective water/HOPG interaction, thus change the free energy. Consequently, the condition of the existence of the micro-pancake state may be destroyed.

We also note that the existence and sizes of gas pancakes and gas bubbles at the liquid/solid interface are affected by the condition of the solid surface, like surface smoothness, cleanness and hydrophobicity. Defects and adsorbents on the surface may divide the solid surface into many small pieces of perfect surface, which limits the coverage of gas pancakes. They may also serve as the nucleation centers for gas bubbles.

One of the key ingredients in the theory of de Gennes and coworkers is the introduction of interactions at the microscopic scale. The equilibrium thickness of a gas pancake corresponds to the minimum of the free energy depicted by eq 4 with the constraint of the total gas molecular number. Consequently, the thickness of the gas pancakes should be within the force range of the long-range interactions at the microscopic scale, whereas the coverage of each pancake can be large. In contrast, the theory can only predict the existence of a non-zero contact angle for possible gas bubbles on a solid surface or on a gas film and the free energy of the possible gas bubbles is not completely covered by the theory. Consequently, the existence of possible stable gas bubbles cannot be completely determined by the present theory. In the macroscopic case, the existence of a non-zero contact angle leads to the state of a stable gas bubble. Considering that macroscopic theories may not be applicable at the nanoscale, gas bubbles on the nanometer scale may also be stable although the lifetime predicted by macroscopic theory is very short. It should be noted that although we cannot resolve the debate on the existence of stable nanoscale gases at the liquid/solid interface, our study shows that both gas bubbles and gas pancakes (or films) with a finite thickness are possible states and there is a new state consisting of gas bubble(s) coexisting with a gas film. These results enrich our current knowledge of gas nucleation and should be helpful for further studies in this direction.

**Conclusions**

In summary, we have analyzed the problem of gas nucleation at the liquid/solid interface by introducing interactions at the microscopic scale. It is found that there are three possible gas states at the liquid/solid interface: complete "wetting", partial "wetting" and pseudopartial "wetting", by analogy with the possible states of liquid nucleation. Our predictions will contribute to a unified picture of gas nucleation at the liquid/solid interface. They would also be helpful in many practical applications and understandings, including the control of the stability of oil/water emulsions,[46,47] detergent-free cleaning,[48] designing of biosensors and biochips,[49] and some important biological processes such as fast protein folding and assembly.[21,50,51]

**Acknowledgment.** The authors thank Professor Xudong Xiao for his discussion, Vassili Yaminsky, Barry Ninham and Scott Edwards for commenting on an early draft of this paper, and Ding Li for his help on the analysis of the density profile of gas on a solid surface from a classical theory. This research is supported in part by grants from Chinese Academy of Sciences, National Science Foundation through projects No.10474109 and 10674146.